\documentclass{elsarticle}

\usepackage{lineno,hyperref}

\journal{arXiv.org}

\biboptions{numbers,sort&compress}

\usepackage{amsmath,amssymb}
\usepackage{mathrsfs}
\usepackage{graphicx}
\usepackage{enumerate}
\usepackage{extarrows}
\usepackage{xcolor}
\usepackage{pdflscape}
\usepackage[version=4]{mhchem}
\usepackage{multirow}
\usepackage{enumerate}

\usepackage[margin=2.5cm]{geometry}

\begin{document}

\begin{frontmatter}

\title{On the relation of the COVID-19 reproduction number to the explosive timescales: the case of Italy}

\author{Dimitris G. Patsatzis\corref{mycorrespondingauthor}}
\address{School of Chemical Engineering, National Technical University of Athens, 15780 Athens, Greece}
\cortext[mycorrespondingauthor]{Corresponding author}
\ead{dpatsatzis@mail.ntua.gr}

%


\begin{abstract}
A great issue of discussion of an infectious disease is its basic reproduction number $R_0$, which provides an estimation of the contagiousness of the disease.~When $R_0>1$, the disease spread will potentially lead to an outbreak, such that of the ongoing COVID-19 pandemics.~During the evolution of an outbreak, various non-pharmaceutical interventions are employed, the impact of which is frequently assessed by the reduction that they introduce to the effective reproduction number $R_t$; reduction below 1 is an indication of eventual dying out of the disease spread.~Motivated by the fact that $R_0$ essentially expresses the stability of the disease-free equilibrium, in this work, $R_t$ was examined in the view of timescale analysis.~It was shown that during the evolution of the COVID-19 outbreak in Italy, when various interventions were in place, $R_t$ had a clear relation with the explosive timescale characterizing the dynamics of the outbreak.~In particular, it is shown that the existence of an explosive timescale during the progression of the epidemics implies $R_t>1$, while its absence implies $R_t<1$.~In addition, as this timescale converges/diverges with the immediately slowest one, $R_t$ approaches to/withdraws from its threshold value 1.~These results suggest that timescale analysis can be utilized for the assessment of the impact of various interventions, since it reflects the insight provided by the effective reproduction number, without being hindered by the selection of the population model, nor the parameter estimation process followed for model calibration.
\end{abstract}


\end{frontmatter}


\section{Introduction}

As of March 11, 2020, the novel coronavirus disease (COVID-19) was declared a pandemic by World Health Organization (WHO) \citep{WHO_pan}.~By January 15, 2021, the COVID-19 pandemics has been spread to more than 219 countries and territories, reporting more than 93 million infected cases and 2 million deaths \citep{Worldometer}.

A great issue of discussion of the COVID-19 pandemics is its basic reproduction number, estimations for which were provided in numerous early studies; see Refs within \citep{liu2020reproductive}.~The basic reproduction number, $R_0$, is the average number of secondary infections produced by an infectious individual in a population where everyone is considered susceptible \citep{delamater2019complexity,diekmann1990definition}.~Being dependent on human behavior and the biological characteristics of the pathogen, $R_0$ provides an estimation of the contagiousness of the infectious disease \citep{delamater2019complexity} and serves as a threshold parameter; when $R_0>1$ the infected increase exponentially, leading to a disease outbreak, while when $R_0<1$ the disease spread dies out \citep{delamater2019complexity,diekmann1990definition}.

For the control of the COVID-19 outbreak, various interventions are employed aiming to ``flatten" the curve of the epidemics.~Since $R_0$ is constant in time, it cannot monitor the effect of the undertaken measures; instead, the time-varying \emph{effective} reproduction number $R_t$ is utilized, that estimates the secondary infections produced by an infectious individual during the course of an outbreak, thus, in a population where not everyone is considered susceptible.~As a result, during the evolution of the epidemics, the undertaken control measures affect $R_t$, since they influence (i) the duration of contagiousness, (ii) the likelihood of infection per contact and (iii) the contact rate of the infection \citep{delamater2019complexity,viceconte2020covid}.~The impact of various interventions (case isolation, contact tracing, travel restrictions, etc.) in $R_t$ has been assessed in a number of studies to provide guidelines in decision-making policies \citep{hellewell2020feasibility,kucharski2020early,pan2020association,cowling2020impact,tang2020estimation,tang2020updated}. 


The use of $R_0$ as a threshold parameter is related to the stability of the disease-free equilibrium (DFE) of the epidemiological model under consideration \citep{diekmann1990definition,van2002,delamater2019complexity}, which is locally assessed by the existence of positive eigenvalues.~During the evolution of the system, the local dynamics is characterized by timescales of dissipative/explosive nature - associated with positive/negative eigenvalues - the action of which tends to drive the system towards to/away from equilibrium \citep{lam1989understanding,maas1992simplifying}.~Timescale analysis has been frequently employed to address the dynamical properties of systems arising from reactive flows \citep{manias2016mechanism,tingas2015autoignition}, systems biology \citep{kourdis2010physical,patsatzis2019new}, pharmacokinetics \citep{patsatzis2016asymptotic}, etc, but, to my knowledge, it hasn't been widely applied to population dynamics.

Motivated by the fact that $R_0$ mathematically expresses the stability of the DFE; i.e., the existence of positive eigenvalues at day zero, here the relation of $R_t$ to the explosive timescales (positive eigenvalues) during the course of COVID-19 outbreak in Italy was investigated.~It is shown that the existence of an explosive timescale implies $R_t>1$, while its absence implies $R_t<1$.~In addition, as this timescale converges/diverges with the immediately slowest one, $R_t$ was shown to approaches to/withdraws from its threshold value 1.~Finally, by performing the analysis in 4 different population dynamics models, it is demonstrated that timescale analysis is a robust methodology to monitor the progression of the epidemics, since it directly reflects the variations in $R_t$, without being hindered by the complexity of the selected model.

\section{Materials and Methods}

Compartmental modeling is widely used for the analysis of various infectious diseases \citep{gao2008pulse,canini2011population,esteva2003coexistence,stegeman1999quantification}, among which COVID-19 pandemics \citep{kucharski2020early,tang2020estimation,giordano2020modelling,russo2020}.~Four population dynamics compartmental models in the framework of the SIR model \citep{kermack1927contribution} were analyzed here.~The effective reproduction number $R_t$ was calculated on the basis of these models and conclusions were drawn on its relation to the timescales characterizing the dynamics of each model.

The four compartmental models are presented in Section~\ref{ss:models}, followed by the parameter estimation process considered for their calibration against the data of Italy in Section~\ref{ss:calib}.~The methodology to calculate the effective reproduction number $R_t$ and the timescales $\tau_i$ on the basis of each model is presented in Sections~\ref{ss:Rt} and \ref{ss:tmscls}, respectively.

\subsection{The population dynamics models}
\label{ss:models}

The SIR model formulates the transmission of an infectious disease among three population groups, namely the \textit{susceptible}, \textit{infected} and \textit{recovered} individuals \citep{kermack1927contribution}.~In this framework, four population dynamics models were considered, the SIRD, SEIRD, SEInsRD and SIDARTHE models, the governing equations of which can be written in the ODE form:
\begin{equation}
\dfrac{d}{dt} \mathbf{y} = \mathbf{g}(\mathbf{y})
\label{eq:VF}
\end{equation}
where $\mathbf{y}$ is the N-dim. column \textit{state vector}, which includes the fraction of each population group over the total population and $\mathbf{g}(\mathbf{y})$ is the N-dim. column \textit{vector field}, which incorporates the transition rates from one population group to another.

The simplest compartmental model to capture COVID-19 pandemics is the SIRD model, which essentially is the SIR model with the addition of a compartment accounting for the \textit{dead} individuals.~Denoting $S$, $I$, $R$ and $D$  the fraction of susceptible, infected, recovered and dead individuals respectively, over the total population $N$, the SIRD model is written in the form of Eq.~\eqref{eq:VF} as:
\begin{equation}
\dfrac{d}{dt} \begin{bmatrix} S \\ I \\ R \\ D \end{bmatrix}= \begin{bmatrix} - \beta SI \\  \beta SI - (\gamma+\mu) I \\ \gamma I \\ \mu I \end{bmatrix} 
\label{eq:SIRD}
\end{equation}
where $\beta$ is the transmission ratio, $\gamma$ the recovery ratio, which also expresses the inverse of the infection period of the disease, and $\mu$ the fatality ratio.
 
A more realistic assumption for COVID-19 infection is the existence of an incubation (latency) period, during which an individual is infected but yet not infectious \citep{lauer2020incubation,li2020early}.~Such an assumption can be incorporated in the SIRD model with the addition of a compartment accounting for \textit{exposed} individuals.~Denoting $E$ their fraction over the total population, the resulting SEIRD model is written in the form of Eq.~\eqref{eq:VF} as:
\begin{equation}
\dfrac{d}{dt} \begin{bmatrix} S \\ E \\ I \\ R \\ D \end{bmatrix}= \begin{bmatrix} - \beta SI \\  \beta SI - \sigma E \\ \sigma E - (\gamma+\mu) I \\ \gamma I \\ \mu I \end{bmatrix} 
\label{eq:SEIRD}
\end{equation}
where $\sigma$ is the transition ratio from exposed to infected individuals, expressing the inverse of the incubation period of the disease.

In addition, it has been shown that the COVID-19 infected individuals have symptoms of different severity, varying from mild to severe \citep{surveillances2020epidemiological}.~Since the severely infected individuals are in need of immediate health care, a more biologically realistic assumption for COVID-19 infection is the distinction between \textit{normally infected} and \textit{severely infected} individuals.~Such an assumption can be incorporated in the SEIRD model, by dividing the infected compartment in two sub-compartments.~Denoting $IN$ and $IS$ the fraction of normally and severely infected individuals over the total population, the resulting SEInsRD model in the form of Eq.~\eqref{eq:VF} reads:
\begin{equation}
\dfrac{d}{dt} \begin{bmatrix} S \\ E \\ IN \\ IS \\ R \\ D \end{bmatrix}= \begin{bmatrix} - \beta_N S.IN - \beta_S S.IS - \mu_{TP} S \\  \beta_N S.IN + \beta_S S.IS -\sigma E - \mu_{TP} E \\ (1-ss) \sigma E - \gamma IN - \mu_N IN  \\ ss \sigma E - \gamma IS - \mu_S IS \\ \gamma (IN+IS) - \mu_{TP} R \\ \mu_N IN + \mu_S IS \end{bmatrix}
\label{eq:SEInsRD}
\end{equation}
where the subscripts $N$ and $S$ indicate the normally and severely infected transmission $\beta$ and fatality $\mu$ ratios, $ss$ denotes the fraction of severely over normally infected individuals and $\mu_{TP}$ is the physiological death ratio.

Finally, a more detailed compartmental model was considered, accounting for susceptible ($S$), asymptomatic detected and undetected infected ($I$ and $D$), symptomatic detected and undetected infected ($A$ and $R$), severely symptomatic ($T$), healed ($H$) and extinct ($E$) individuals, namely the SIDARTHE model \cite{giordano2020modelling}.~Here, the SIDARTHE model was considered for validation purposes and thus, only a brief description of the model is provided in \ref{app:Rt}; details can be found in \cite{giordano2020modelling}.~Note that the SIDARTHE model is also written in the form of Eq.~\eqref{eq:VF}; see Eq.~\eqref{eq:SIDARTHE1} and Eqs.~(1-8) in \textit{Methods} section in \cite{giordano2020modelling}.

\subsection{Model calibration}
\label{ss:calib}

In this study only the SEIRD and SEInsRD models in Eqs.~(\ref{eq:SEIRD}, \ref{eq:SEInsRD}) respectively, were calibrated, since (i) the relation of $R_t$ with the timescales can be reached analytically on the basis of the SIRD model in Eq.~\eqref{eq:SIRD} and (ii) the parameter values of the SIDARTHE model are provided in \citep{giordano2020modelling}.

The SEIRD and SEInsRD models in Eqs.~(\ref{eq:SEIRD}, \ref{eq:SEInsRD}) were calibrated to the daily reported data of infected, recovered and dead individuals in Italy, as reported by John Hopkins database \cite{JH_db}.~The parameter estimation process was performed in a weekly basis accounting for the data from February 26 (week 0) to September 30 (week 30).~February 26 was selected as starting day in order to minimize early data distortion, since more than 400 infected individuals were reported at that date.~Initially, given the reported fraction of infected, recovered and dead population groups at week 0 - and the susceptible one, through conservation of the total population - the fraction of the exposed, normally infected and severely infected population groups was estimated.~In the following, a parameter estimation process was performed in a weekly basis, given these 3 reported data sets, through a genetic algorithm provided by the open-source COPASI software \citep{runarsson2000stochastic,hoops2006copasi}.~The initial conditions at day 0 of each week were the predicted values at day 7 of the previous week, in order to preserve continuity in the solution.~The resulting parameter sets are depicted for SEIRD and SEInsRD models in Fig.~\ref{fig:Params_SEIRD+SEInsRD} of \ref{app:ModParam}.


\begin{figure}[!h]
\centering
\includegraphics[scale=0.35]{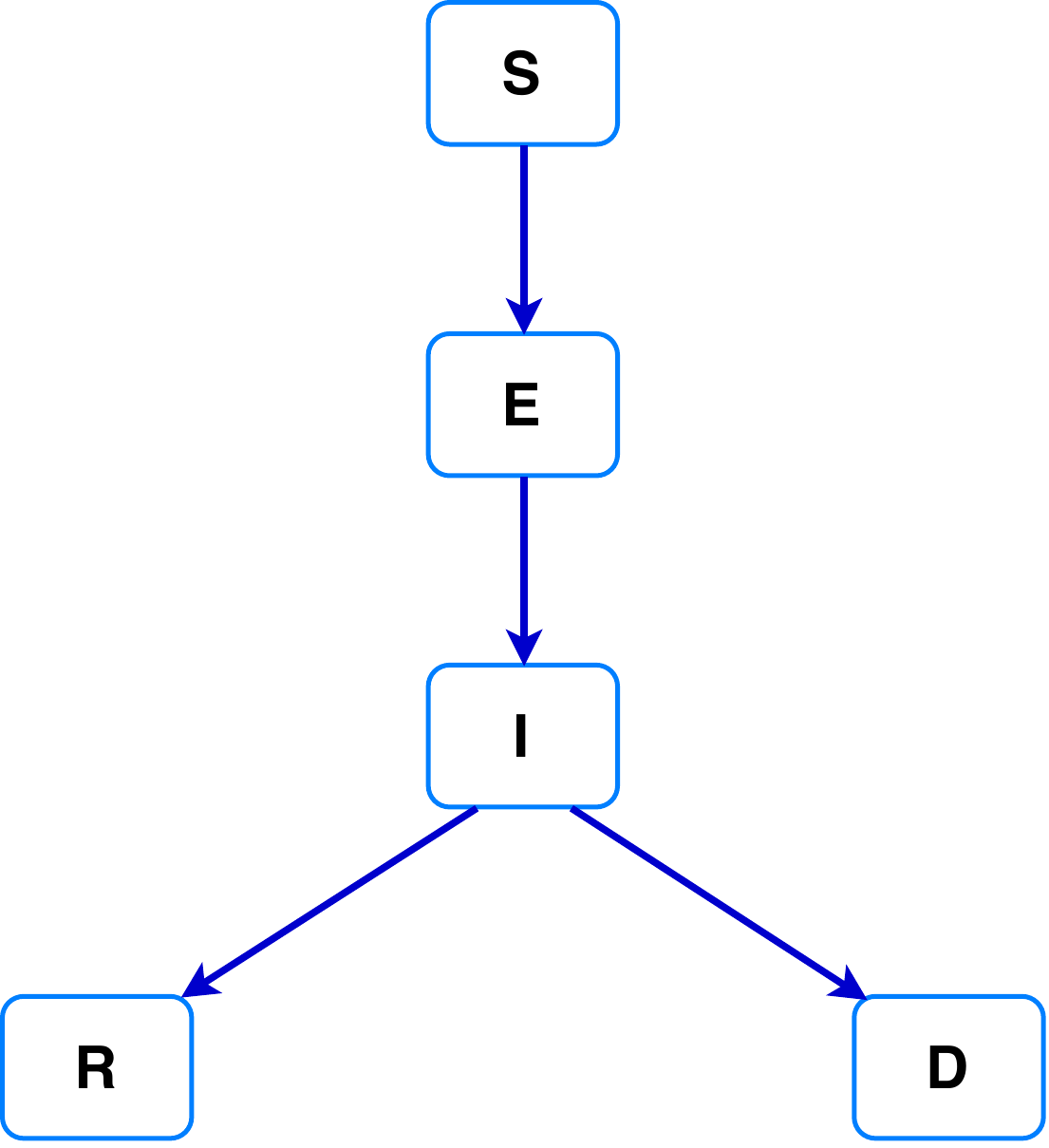} \hspace{3pt} \includegraphics[scale=0.22]{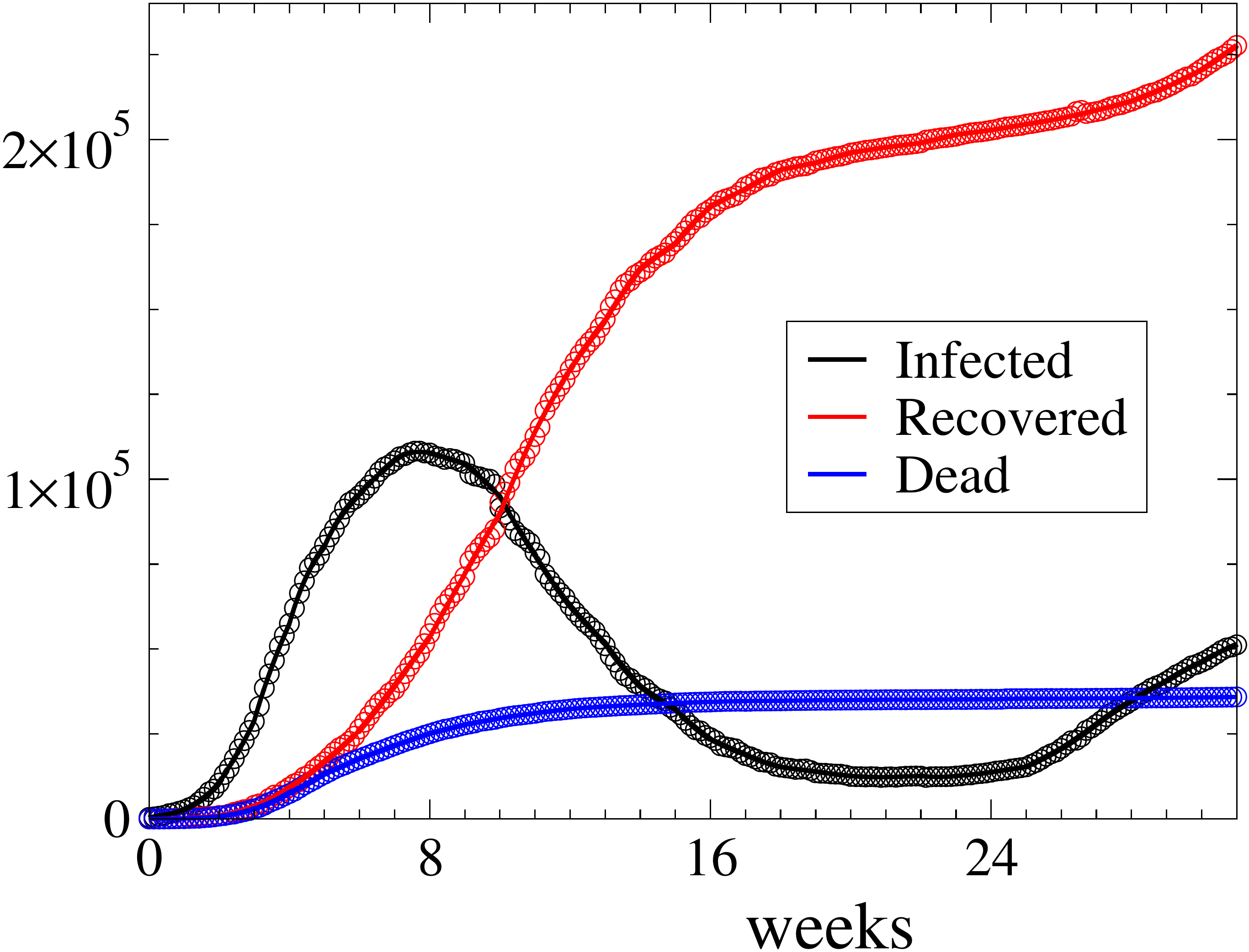}  \hspace{3pt} \includegraphics[scale=0.22]{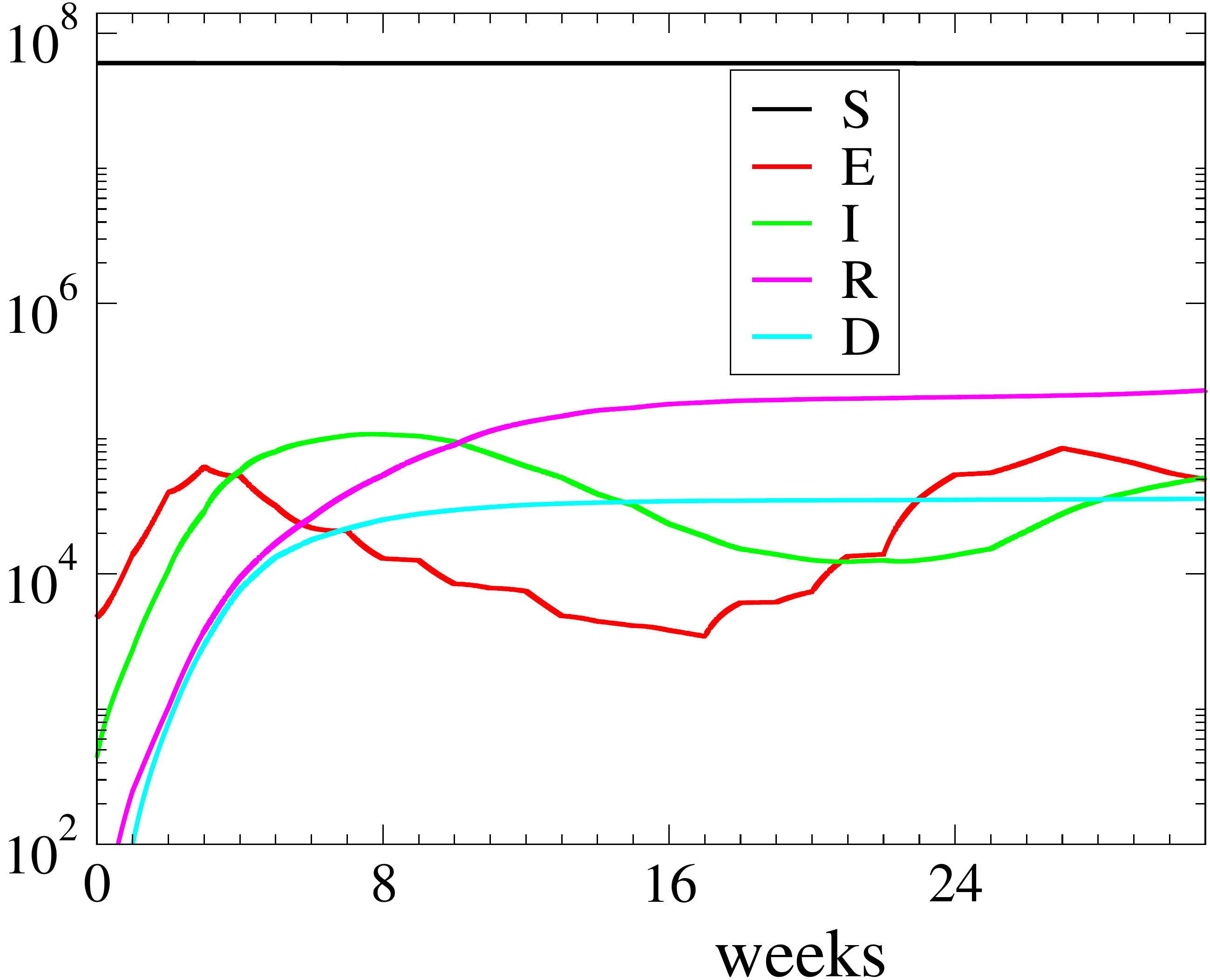}
\caption{The SEIRD model (left), its fitting against the reported data for Italy in circles (middle) and the profiles of all the population groups (right).~The parameter estimation process was performed in a weekly basis from the 26th of February (week 0) to the 30th of September (week 30), accounting for the reported data of infected, recovered and dead individuals.}
\label{fig:Fit_SEIRD}
\end{figure}

\begin{figure}[!h]
\centering
\includegraphics[scale=0.35]{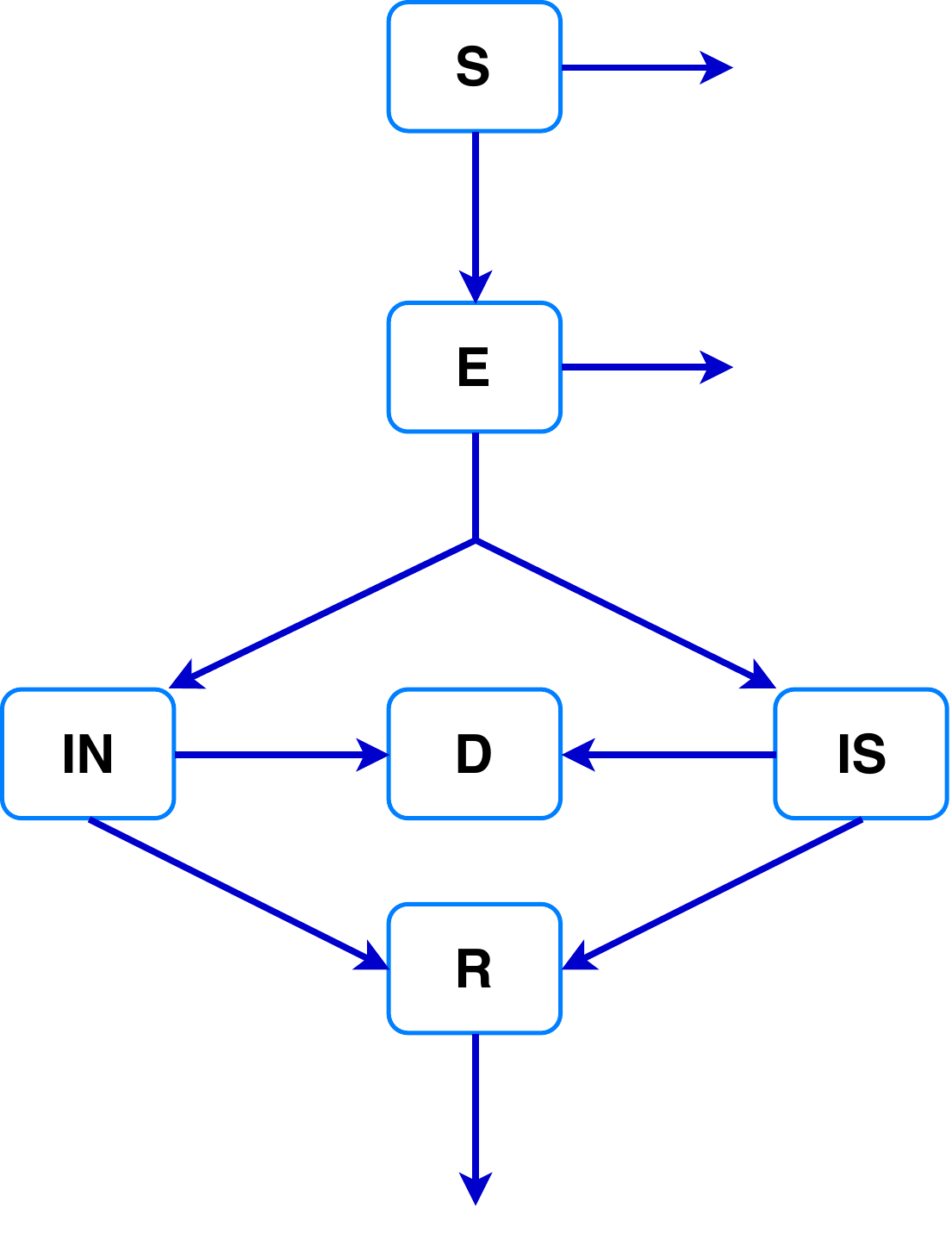} \hspace{3pt} \includegraphics[scale=0.22]{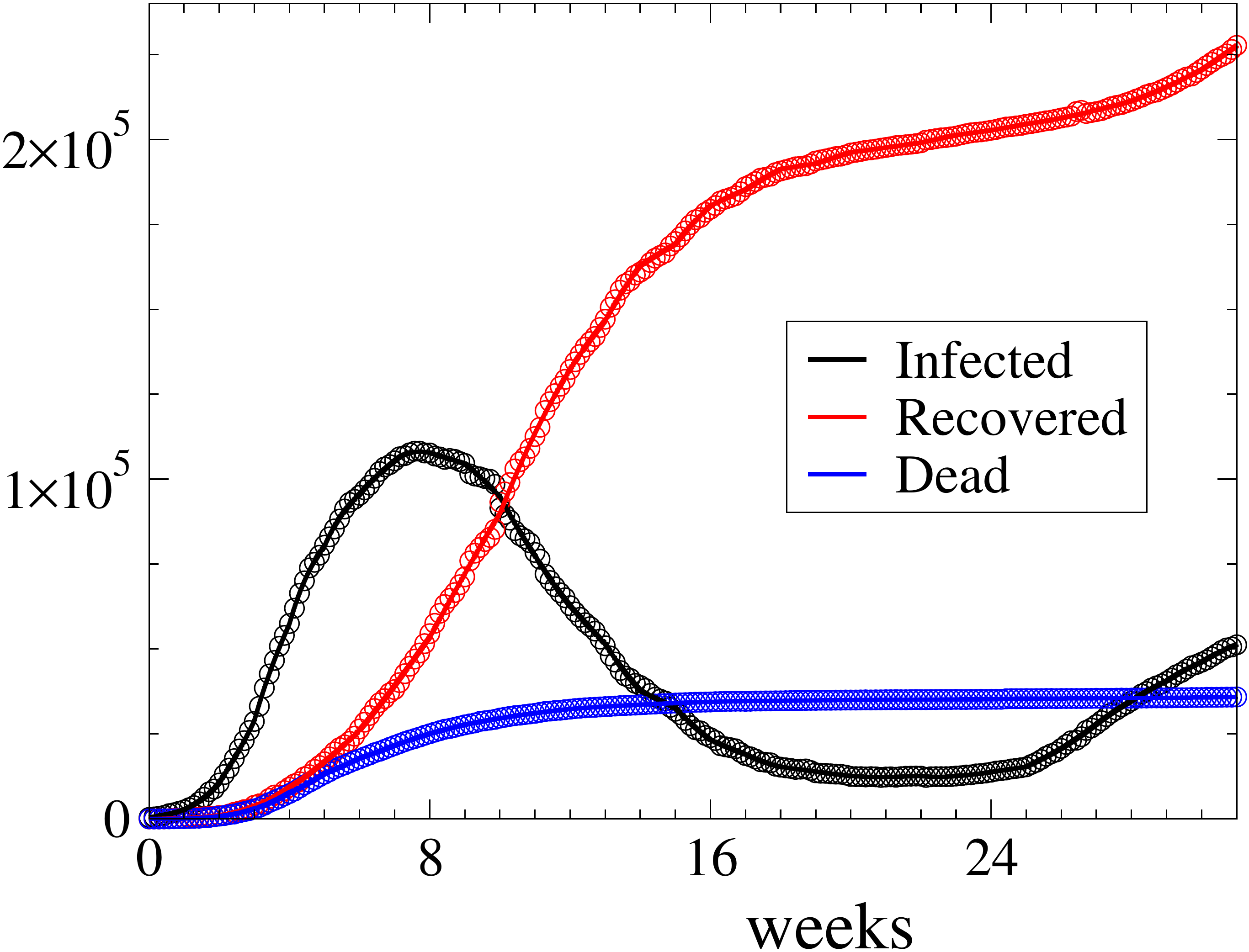}  \hspace{3pt} \includegraphics[scale=0.22]{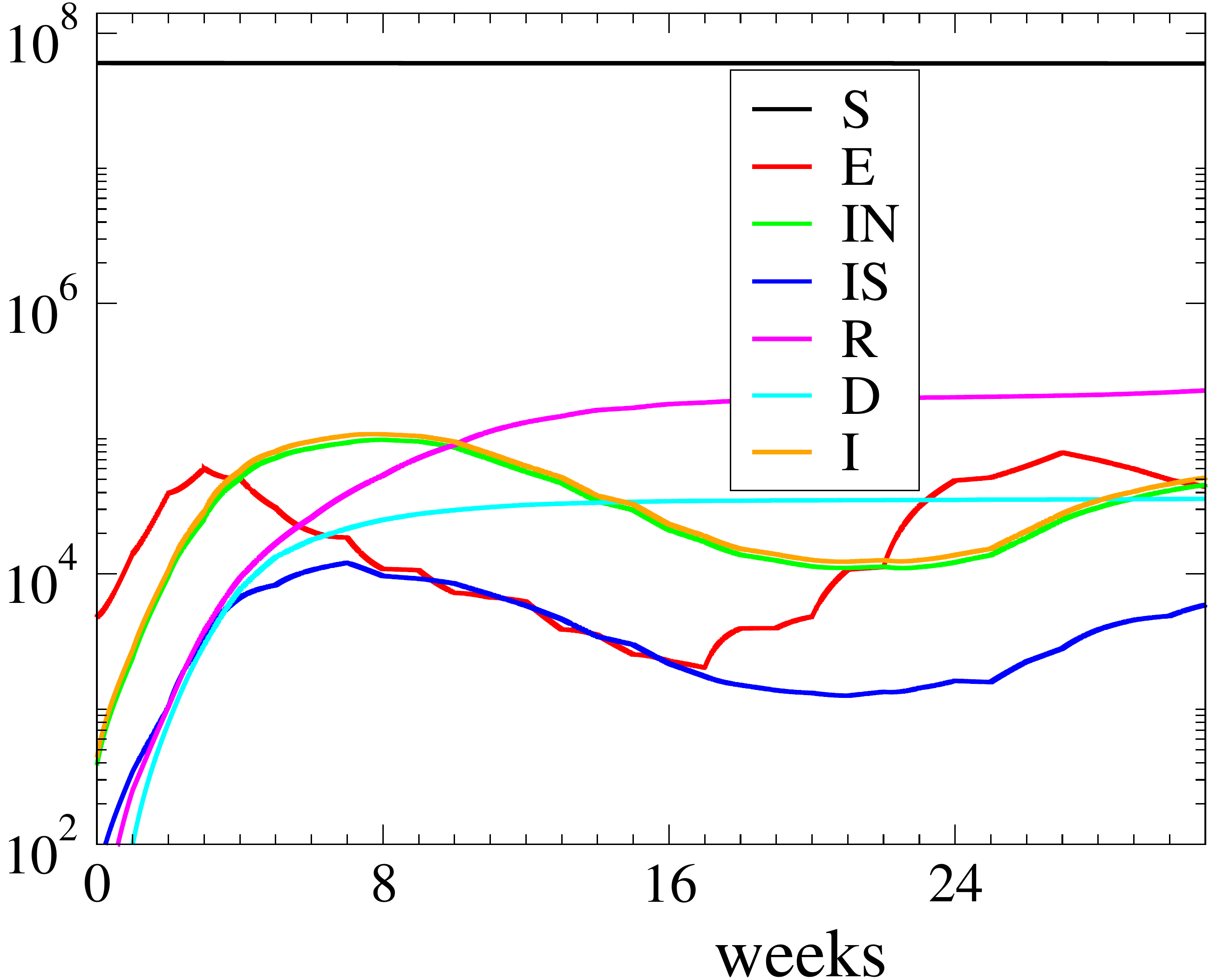}
\caption{The SEInsRD model (left), its fitting against the reported data for Italy in circles (middle) and the profiles of all the population groups (right).~The parameter estimation process was performed in a weekly basis from the 26th of February (week 0) to the 30th of September (week 30), accounting for the reported data of infected, recovered and dead individuals.}
\label{fig:Fit_SEInsRD}
\end{figure}

A schematic representation of the SEIRD and SEInsRD models is provided in the left panels of Figs.~\ref{fig:Fit_SEIRD} and \ref{fig:Fit_SEInsRD}, respectively.~The profiles of infected, recovered and dead individuals, resulted from the aforementioned parameter estimation process, are in very good agreement with the reported data, as shown in the middle panels of Figs.~\ref{fig:Fit_SEIRD} and \ref{fig:Fit_SEInsRD} for the SEIRD and SEInsRD models, respectively.~Note that in the case of SEInsRD model, the sum of the normally and severely infected individuals $I=IN+IS$ was fitted against the reported data set of the infected individuals.~The very good agreement of the model parameters to the reported data can be also demonstrated by the $R^2$ values of both fittings shown in Table \ref{tb:ParamRange+Rs+p} of \ref{app:ModParam}, combined with the respective p-values, which in all cases are $p\ll0.05$.~Finally, the profiles of all the population groups are displayed in the right panels of Figs.~\ref{fig:Fit_SEIRD} and \ref{fig:Fit_SEInsRD}, in which great agreement of the population profiles between the two models is reported.

Finally, the SIDARTHE model parameters were directly adopted by \citep{giordano2020modelling}, following a slightly different approach than the one followed here for SEIRD and SEInsRD models.~First, due to availability of data, here the SEIRD and SEInsRD models were calibrated for Italy from February 26 to September 30, while the SIDARTHE model was calibrated for Italy from February 20 to April 5.~Second, here the SEIRD and SEInsRD models were calibrated in a constant, 7-days long, time frame, while the SIDARTHE model was calibrated in varying time frames, depending on the interventions undertaken in Italy \citep{giordano2020modelling}.~Last but not least, the reported data of infected, recovered and dead individuals were considered in our analysis, while in \citep{giordano2020modelling} only the ones of infected and recovered (not fitted to the \textit{healed} compartment of the model) individuals.

\subsection{Estimation of the reproduction number}
\label{ss:Rt}

The basic reproduction number, $R_0$, is a constant biological parameter that provides an estimation of the contagiousness of the infectious disease.~It also serves as a threshold parameter; when $R_0>1$, one infected individual can trigger an outbreak, while when $R_0<1$, the infection will not spread in the population \citep{diekmann1990definition,delamater2019complexity}.

When various non-pharmaceutical interventions (NPI) are in place, the effective reproduction number $R_t$ is utilized, instead of $R_0$, to monitor the reproduction number during the evolution of the outbreak.~$R_t$ provides an estimation of the contagiousness of the infectious disease, during the course of an outbreak, where not every individual is considered susceptible.~Considering that all model parameters are time dependant, we estimated $R_t$ for COVID-19 pandemics in Italy using the Next Generation Matrix (NGM) approach \citep{heffernan2005,van2017,van2002}, which yields in the following expressions for the SIRD, SEIRD, SEInsRD and SIDARTHE models:
\begin{align} 
R^{SIRD}_t & =  \dfrac{\beta}{\gamma+\mu} =R^{SEIRD}_t   \nonumber \\ 
R^{SEInsRD}_t & = \dfrac{\sigma}{\sigma+\mu_{TP}} \left( \dfrac{(1-ss) \beta_N}{\gamma+\mu_N} +\dfrac{ss \beta_S}{\gamma+\mu_S}\right)
\label{eq:Rt} \\
R^{SIDARTHE}_t & = \dfrac{\alpha}{r_1} + \dfrac{\beta \epsilon}{r_1 r_2} + \dfrac{\gamma \zeta}{r_1 r_3} + \dfrac{\delta \theta \zeta}{r_1 r_3 r_4} + \dfrac{\delta \epsilon \eta}{r_1 r_2 r_4} \nonumber
\end{align}
where $\alpha, \beta, \gamma, \delta, \epsilon, \zeta, \eta, \theta$ are model parameters of the SIDARTHE model and $r_1=\epsilon+\lambda+\zeta$, $r_2=\eta+\rho$, $r_3=\kappa+\mu+\theta$ annd $r_4=\nu+\xi$.~Note that the expression of $R_t$ for SIDARTHE model estimated here via the NGM approach is the same with the one derived in \citep{giordano2020modelling}.

A brief discussion on NGM approach is provided in \ref{app:Rt}, along with details on the calculation of $R_t$ on the basis of the four population dynamics models.

\subsection{Calculation of the time scales}
\label{ss:tmscls}

Given a system of ODEs in the matrix form of Eq.~\eqref{eq:VF}, the timescales are calculated as the inverse modulus of the eigenvalues of the N$\times$N Jacobian matrix $\mathbf{J}(\mathbf{y})=\nabla_{\mathbf{y}} \left(\mathbf{g}(\mathbf{y})\right)$ \citep{lam1989understanding,maas1992simplifying}.~The timescales are of dissipative/explosive nature, i.e., the components of the system that generate them tend to drive the system towards to/away from equilibrium, when the respective eigenvalue has negative/positive real part.

When a complex mathematical model in the form of Eq.~\eqref{eq:VF} is encountered, it is usually impossible to calculate  analytic expressions for its eigenvalues and thus its timescales.~This is the case of the SEIRD, SEInsRD and SIDARTHE models, for which the timescales were calculated numerically.~However, in the case of the SIRD model, the non-zero eigenvalues can be calculated analytically as:
\begin{equation}
\lambda_{1,2}=\dfrac{1}{2} \left( X \pm \sqrt{X^2-4Y} \right) \qquad \qquad X=-\gamma-\beta I - \mu+ \beta S \qquad Y=\beta I (\gamma+\mu)
\label{eq:eigenSIRD}
\end{equation}
Therefore, the related timescales are of explosive nature (either real or complex $\lambda_{1,2}$) if and only if:
\begin{equation}
X>0 \Rightarrow \beta (S-I)>\gamma+ \mu \Rightarrow \dfrac{\beta (S-I)}{\gamma+\mu}>1
\label{eq:xx1}
\end{equation}
Equation \eqref{eq:xx1} provides the condition under which the explosive timescales of the SIRD model arise, a feature that will associated in the following section with $R_t$.

\section{Results}

The impact of the undertaken NPIs in COVID-19 pandemics is assessed by the effect that they introduce in the reproduction number \citep{hellewell2020feasibility,kucharski2020early,pan2020association,cowling2020impact,tang2020estimation,tang2020updated}.~Here, we show that the insights provided by the utilization of the effective reproduction number $R_t$ during the progression of the COVID-19 pandemics can be deduced by timescale analysis.~In particular, it is shown that:
\begin{enumerate}[i)]
\item the existence of an explosive timescale during the progression of COVID-19 epidemics implies $R_t>1$, while its absence implies $R_t<1$, and
\item the tendency of this timescale to converge/diverge with the immediately slowest one, implies that $R_t$ tends to approach to/withdraw from its threshold value 1.
\end{enumerate}
These results are reached on the basis of the four population dynamics models discussed in Section~\ref{ss:models}, for the case of Italy.

\subsection{The explosive timescales in relation to the reproduction number}
\label{ss:SIRD+SEIRD}

The first indication on the relation of the explosive timescales to the reproduction number is provided by the analysis of the SIRD model in Eq.~\eqref{eq:SIRD}, that is the simplest model to describe the progression of COVID-19 epidemics.~In contrast to more complicated models, the timescales of the SIRD model can be calculated analytically.~According to Section~\ref{ss:tmscls}, the evolution of the SIRD model is characterized by the action of two timescales $\tau_{1,2}=1/|\lambda_{1,2}|$; the expressions of $\lambda_{1,2}$ were derived in Eq.~\eqref{eq:eigenSIRD}.~Both $\tau_{1,2}$ are of explosive/dissipative nature when the condition in Eq.~\eqref{eq:xx1} holds/is violated.~Given the expression of $R_t$ for SIRD model in Eq.~\eqref{eq:Rt} and that $S-I<S(0)=1$, Eq.~\eqref{eq:xx1} yields:
\begin{equation}
Re(\lambda_{1,2})>0 \Leftrightarrow  X>0 \Leftrightarrow \dfrac{\beta (S-I)}{\gamma+\mu}>1 \Rightarrow  \dfrac{\beta S(0)}{\gamma+\mu}>1 \Rightarrow R_t>1
\label{eq:xx2}
\end{equation}

Equation~\eqref{eq:xx2} shows that the existence of explosive timescales implies $R_t>1$, while their absence implies $R_t<1$.~Note that this outcome, holds true not only for COVID-19 pandemics, but also for any infectious disease, since it was derived by analytical means on the basis of the SIRD model.

Next, the relation of the explosive timescales with $R_t$ was examined using reported data for COVID-19 pandemics in Italy.~The SEIRD model in Eq.~\eqref{eq:SEIRD} was adopted and fitted against the reported data sets of infected, recovered and dead individuals in Italy from February 26 to September 30.~In order to account for the NPIs undertaken, the SEIRD model was calibrated in a weekly basis following the parameter estimation process described in detail in Section~\ref{ss:calib}.~The resulting solution is in great agreement with the reported data, as shown in Fig.~\ref{fig:Fit_SEIRD}.

\begin{figure}[!b]
\centering
\includegraphics[scale=0.25]{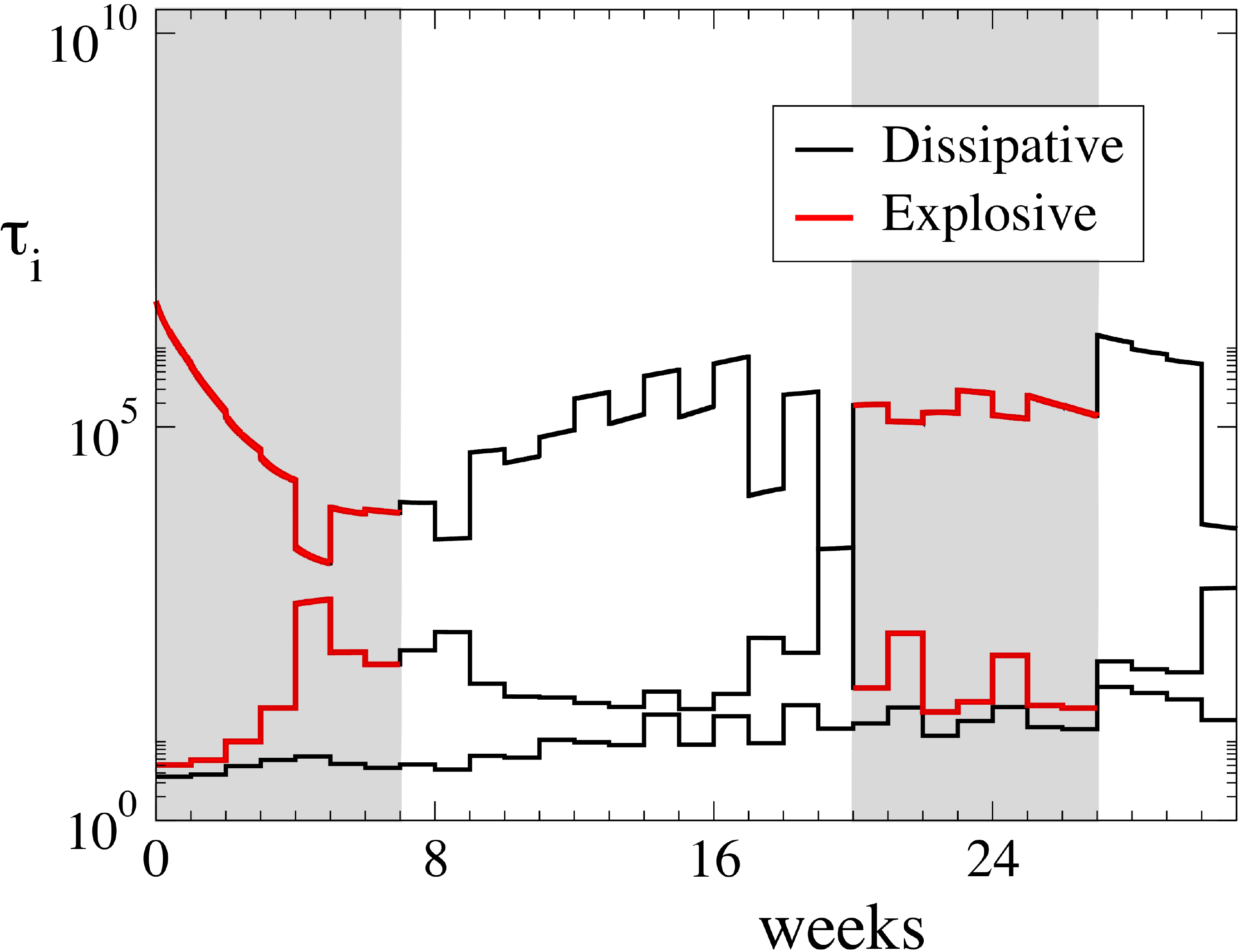} \hspace{25pt} \includegraphics[scale=0.25]{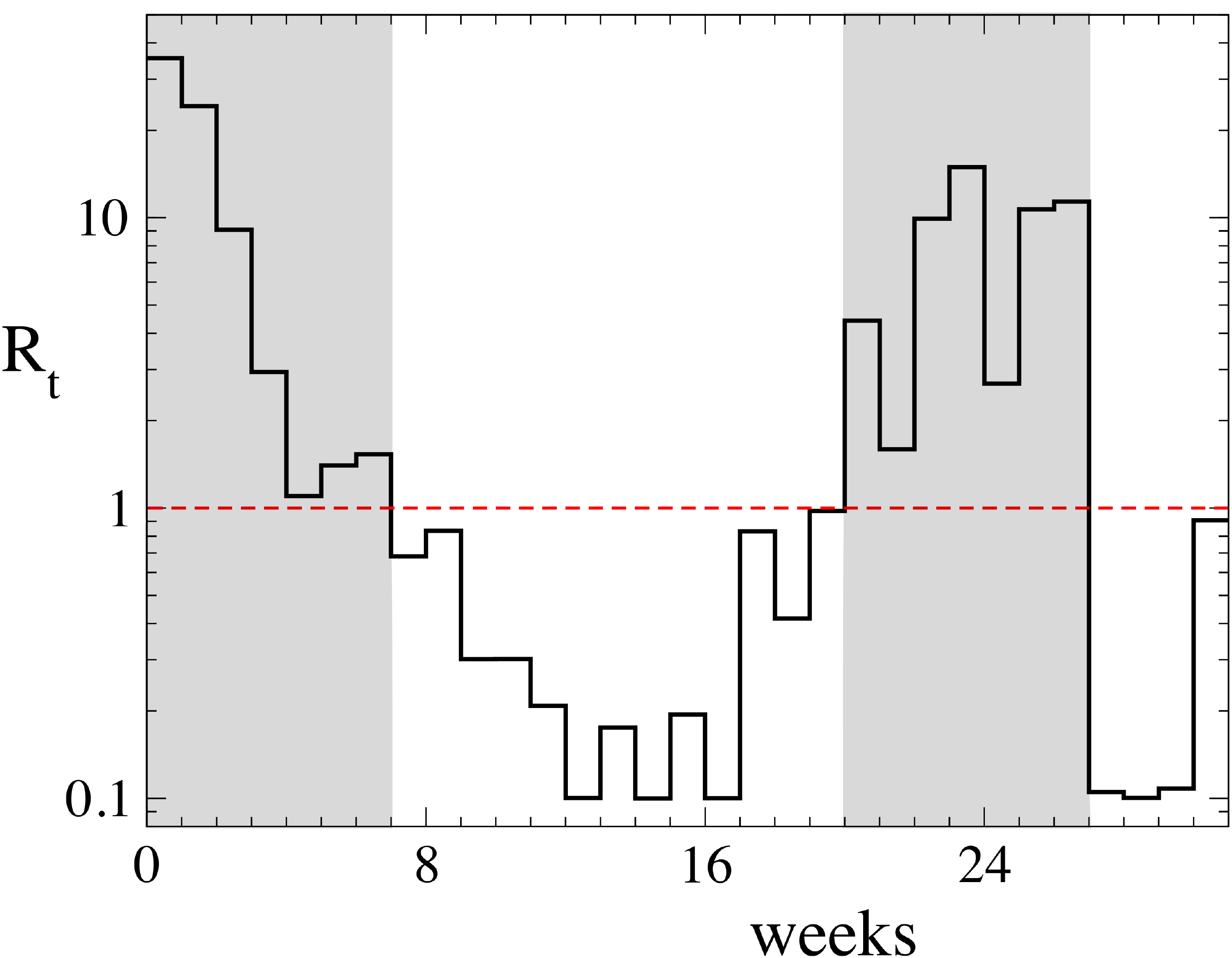}  
\caption{The timescales (left) and the effective reproduction number $R_t$ (right) estimated on the basis of the solution of the SEIRD model shown in Fig.~\ref{fig:Fit_SEIRD}.} 
\label{fig:Tmscls+R0_SEIRD}
\end{figure}

The timescales and $R_t$, estimated on the basis of the SEIRD model in Eq.~\eqref{eq:Rt}, are displayed in Fig.~\ref{fig:Tmscls+R0_SEIRD} from week 0 (starting in Feb. 26) to week 30 (ending in Sep. 30).~As shown in the left panel of Fig.~\ref{fig:Tmscls+R0_SEIRD}, the evolution of the SEIRD model is characterized by three timescales $\tau_{1,2,3}$, the fastest of which, $\tau_{1}$, is always dissipative in nature, while $\tau_{2,3}$ are either dissipative or explosive.~In particular, during weeks 0-6 and 20-26, $\tau_{2,3}$ are explosive, as indicated by the shaded background in Fig.~\ref{fig:Tmscls+R0_SEIRD}.~The values of $R_t$ are depicted in the right panel of Fig.~\ref{fig:Tmscls+R0_SEIRD}, in which the red dashed horizontal line indicates the threshold value $R_t=1$.~As indicated by the shaded background, the time periods when the explosive nature of timescales $\tau_{2,3}$ is reported coincides with the ones that $R_t>1$ (weeks 0-6 and 20-26).~In contrast, when $\tau_{2,3}$ are of dissipative nature, $R_t<1$ (weeks 7-19, 27-30).~Note that the transition from the explosive to the dissipative nature of the timescales $\tau_{2,3}$, and vice-versa, is immediate, since model calibration is performed in a weekly basis.

Comparison of the explosive timescales and $R_t$ in Fig.~\ref{fig:Tmscls+R0_SEIRD} reveals the following trend: as the gap between $\tau_2$ and $\tau_3$ decreases/increases, $R_t$ approaches to/withdraws from its threshold value unity.~This is particularly clear during the first wave of COVID-19 pandemics in Italy (weeks 0-12).~During weeks 0-6, where $\tau_2$ and $\tau_3$ are explosive, their gap tends to decrease, so that $R_t$ decreases, approaching close to unity values.~At week 7, $\tau_2$ and $\tau_3$ become dissipative and $R_t$ attains values below 1.~From this point on and up to week 12, the gap of $\tau_2$ and $\tau_3$ increases, so that $R_t$ continues to decrease, this time withdrawing from its threshold 1.~This behaviour is additionally supported by the fact that during weeks  4-8, 17, 19, 21 and 30, in which $R_t$ attains close to 1 values, the gap between $\tau_2$ and $\tau_3$ is small; to the point where $\tau_2=\tau_3$ in week 19, in which $R_t=0.96$.

\subsection{Robustness}
\label{ss:rob}

In order to demonstrate the robustness of the relation of the explosive timescales to the reproduction number, a more complicated population dynamics model was considered, the SEInsRD model.~The SEInsRD model in Eq.~\eqref{eq:SEInsRD} was adopted and calibrated to the same reported data sets with SEIRD model in Section~\ref{ss:SIRD+SEIRD}, corresponding to infected, recovered and dead individuals in Italy from February 26 to September 30.~Similarly to SEIRD model, the SEInsRD model calibration was performed in a weekly basis following the process described in Section~\ref{ss:calib} and the resulting solution is in great agreement with the reported data, as shown in Fig.~\ref{fig:Fit_SEInsRD}.

\begin{figure}[!h]
\centering
\includegraphics[scale=0.25]{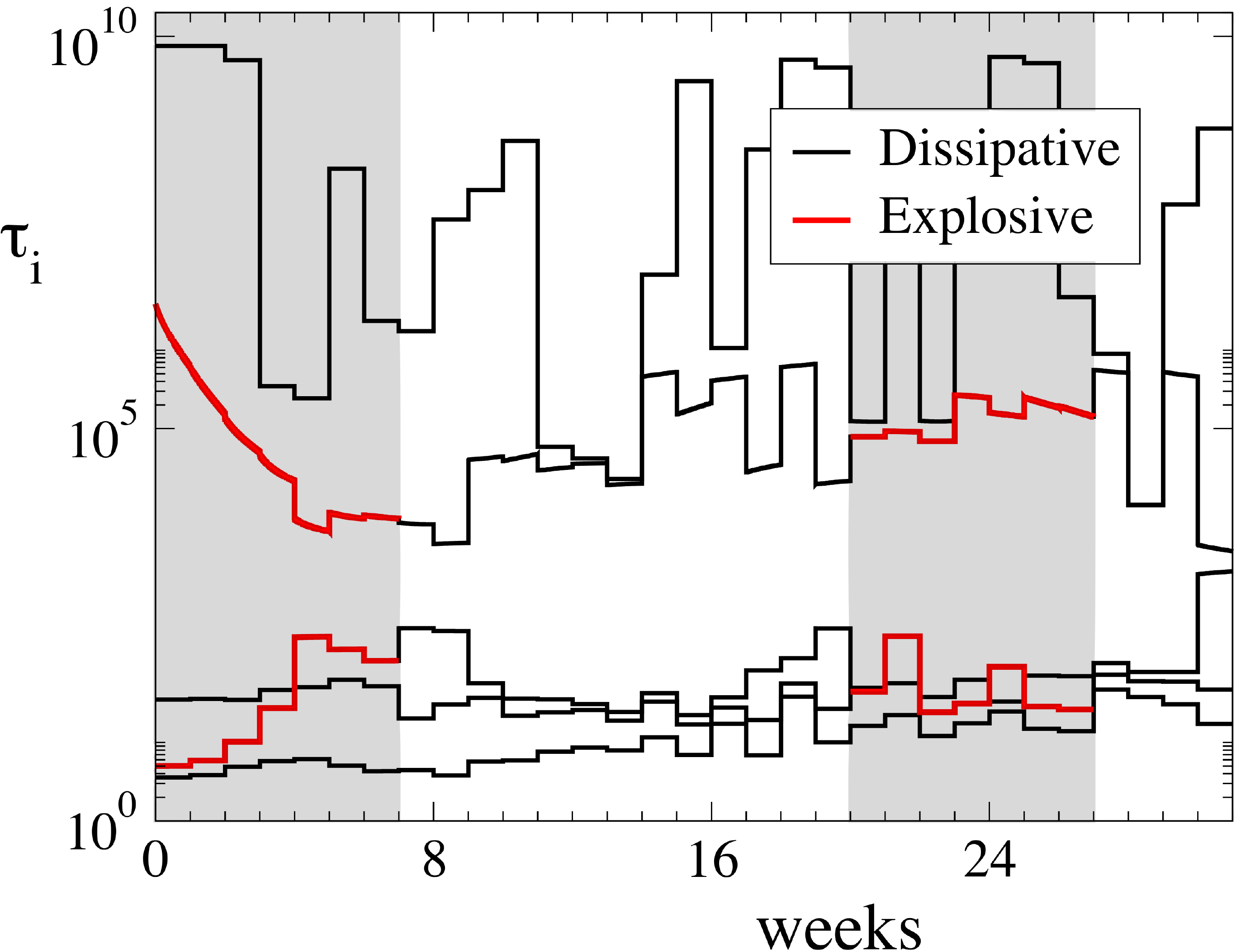} \hspace{25pt} \includegraphics[scale=0.25]{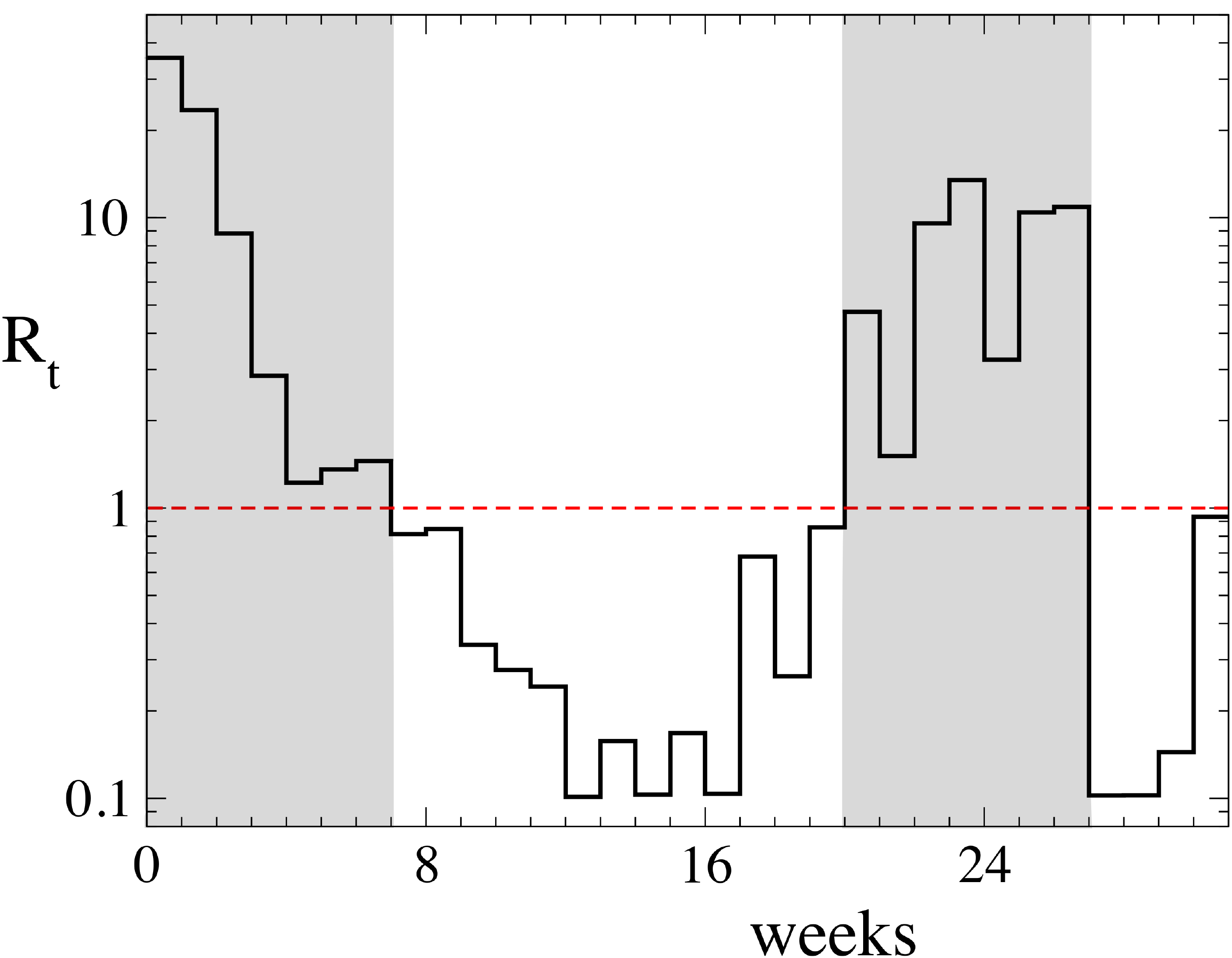}  
\caption{The timescales (left) and the reproduction number $R_t$ (right) calculated on the basis of the solution of the SEInsRD model shown in Fig.~\ref{fig:Fit_SEInsRD}.}
\label{fig:Tmscls+R0_SEInsRD}
\end{figure}

The timescales and $R_t$, estimated on the basis of the SEInsRD model in Eq.~\eqref{eq:Rt}, are displayed in Fig.~\ref{fig:Tmscls+R0_SEInsRD} from week 0 (starting in Feb. 26) to week 30 (ending in Sep. 30).~As shown in the left panel of Fig.~\ref{fig:Tmscls+R0_SEInsRD}, the evolution of SEInsRD model is characterized by 5 timescales: three of which are always of dissipative nature and the remaining ones are either explosive or dissipative; denoting $\tau_{exp,f}$ the fast explosive timescale and $\tau_{exp,s}$ the slow one.~In particular, during weeks 0-6 and 20-26, $\tau_{exp,f}$ and $\tau_{exp,s}$ are explosive, as indicated by the shaded background in Fig.~\ref{fig:Tmscls+R0_SEInsRD}.~The right panel of Fig.~\ref{fig:Tmscls+R0_SEInsRD} displays the values of $R_t$ in comparison to the threshold value $R_t=1$ indicated by the red dashed horizontal line.~Similarly to the SEIRD model, it is shown by the shaded background that $R_t>1$ when the $\tau_{exp,f}$ and $\tau_{exp,s}$ are explosive (weeks 0-6 and 20-26), while $R_t<1$ when they lose this character and become dissipative (weeks 7-19 and 27-30).~In addition, the trend of increasing/decreasing gap of $\tau_{exp,f}$ and $\tau_{exp,s}$ is again reflected in $R_t$ approaching to/withdrawing from its threshold value 1.~In particular, it is shown that the closer the values of $R_t$ to 1, (weeks  4-8, 17, 19, 21 and 30), the smaller the gap between $\tau_{exp,f}$ and $\tau_{exp,s}$; to the point where $R_t=0.97$ in week 30, in which $\tau_{exp,f} \approx \tau_{exp,s}$.~In summary, the qualitative results on the relation of the explosive timescales to $R_t$ are maintained on the basis of the SEInsRD model.


\subsection{Validation}

In order to validate the qualitative results, reached on the basis of the SEIRD and SEInsRD models, regarding to the relation of the explosive timescales to $R_t$, a more complicated SIDARTHE model was considered \citep{giordano2020modelling}, as briefly discussed in Section~\ref{ss:models}.

\begin{figure}[!t]
\centering
\includegraphics[scale=0.25]{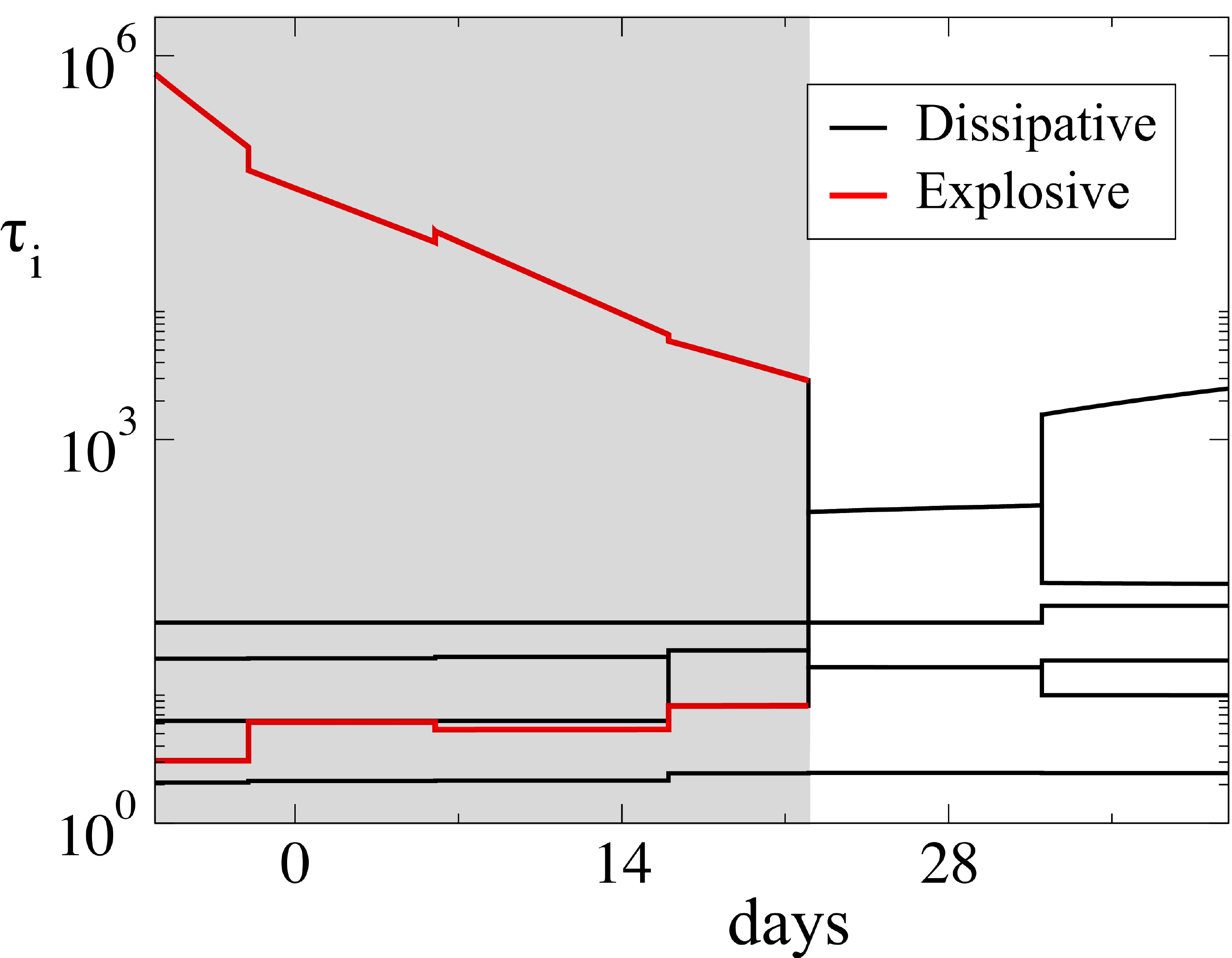} \hspace{25pt} \includegraphics[scale=0.25]{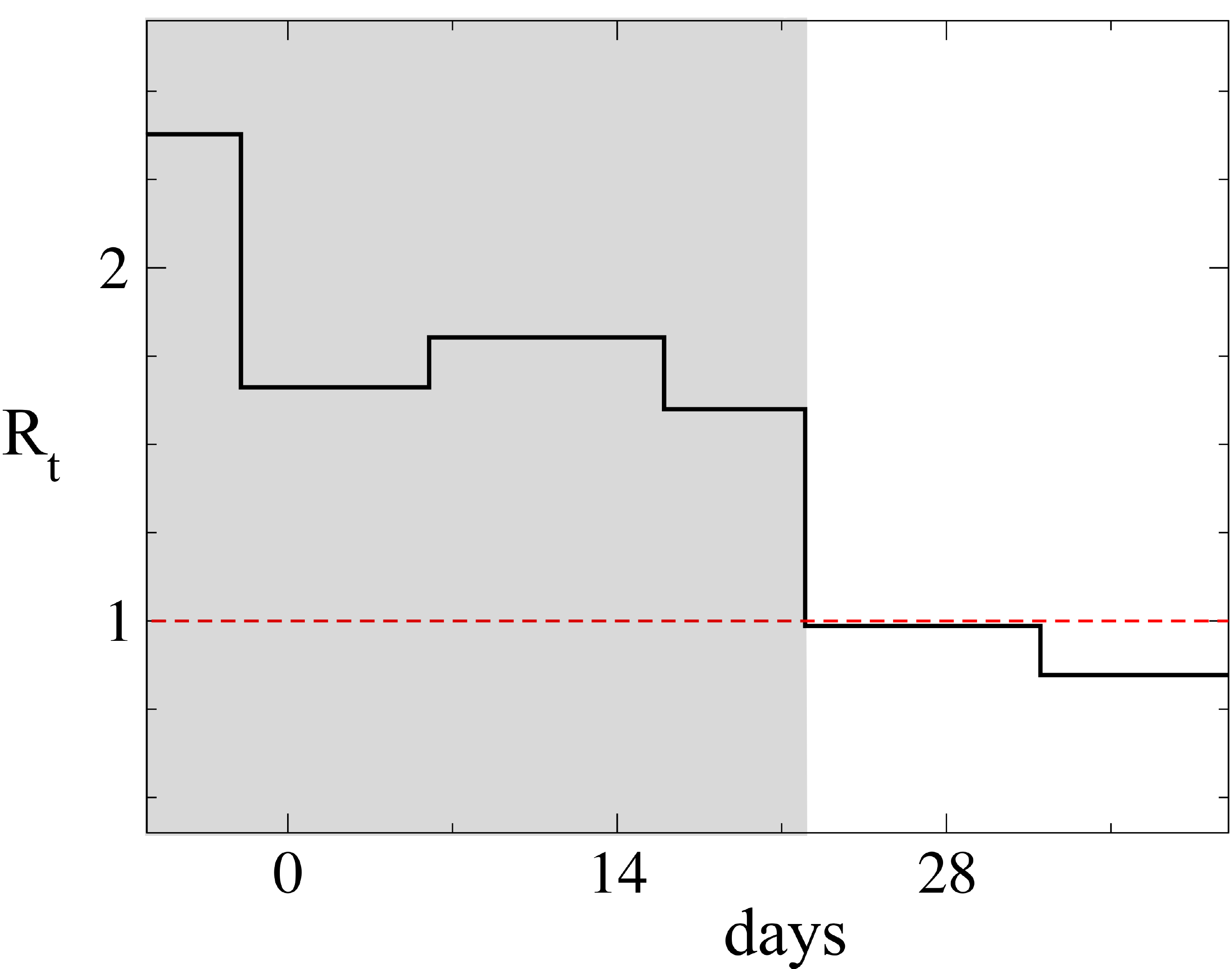}
\caption{The timescales and reproduction number R$_t$ calculated on the basis of SIDARTHE model, that was calibrated for Italy data in \citep{giordano2020modelling}.}
\label{fig:Tmscls+R0_SIDARTHE}
\end{figure}

The profiles of the population groups accounted for in the SIDARTHE model were reproduced, adopting the model parameters in \citep{giordano2020modelling}.~On the basis of the SIDARTHE solution, the timescales were calculated and $R_t$ was estimated according to the expression in Eq.~\eqref{eq:Rt}.~The resulting values are displayed in Fig.~\ref{fig:Tmscls+R0_SIDARTHE} starting from day -6 (Feb 20) and ending in day 40 (Apr 5); day 0 was chosen to be Feb 26 for comparison with Figs.~\ref{fig:Tmscls+R0_SEIRD} and \ref{fig:Tmscls+R0_SEInsRD}.~As shown in the left panel of Fig.~\ref{fig:Tmscls+R0_SIDARTHE}, the evolution of SIDARTHE model is characterized by six timescales, four among which are always dissipative in nature, while the remaining two are either dissipative or explosive; denoted as $\tau_{exp,f}$ and $\tau_{exp,s}$.~In particular, $\tau_{exp,f}$ and $\tau_{exp,s}$ are explosive from day -6 to day 22, as indicated by the shaded background in Fig.~\ref{fig:Tmscls+R0_SIDARTHE}.~The values of $R_t$ are depicted in the right panel of Fig.~\ref{fig:Tmscls+R0_SIDARTHE}, in which the red dashed horizontal line indicates the threshold value $R_t=1$.~As indicated by the shaded background, the explosive nature of timescales $\tau_{exp,f}$ and $\tau_{exp,s}$ implies $R_t>1$ (days -6-22), while losing such a nature and becoming dissipative, implies $R_t<1$ (days 23-40).~In addition, it is shown that as the gap between $\tau_{exp,f}$ and $\tau_{exp,s}$ becomes smaller, $R_t$ approaches its threshold value unity, to the point when $R_t=0.986$ during days 23-33, in which $\tau_{exp,f}=\tau_{exp,s}$.

It should be noted here that, the evolution of timescales $\tau_{exp,f}$ and $\tau_{exp,s}$ and the reproduction number $R_t$, calculated on the basis of SIDARTHE model, is different in comparison to those calculated on the basis of SEIRD and SEInsRD models.~In particular, on the basis of the SIDARTHE model, the timescales are explosive in nature and $R_t>1$ until to day 22, while on the basis of the SEIRD and SEInsRD models until day 42.~Despite this being a major difference, that originates from differences in model calibration as discussed in Section~\ref{ss:calib}, the relation of the explosive timescales to $R_t$, deduced on the basis of SEIRD and SEInsRD models, is validated by the analysis with the SIDARTHE model.

\section{Conclusions}

The progression of an infectious disease spread like COVID-19 pandemics is frequently examined by population dynamics models \citep{gao2008pulse,canini2011population,esteva2003coexistence,stegeman1999quantification,kucharski2020early,tang2020estimation,giordano2020modelling,russo2020}.~Their evolution as dynamical systems is characterized by timescales that are either of dissipative or explosive nature; i.e., their action tends to drive the system either towards to or away from equilibirum \citep{lam1989understanding,maas1992simplifying}.~The basic reproduction number $R_0$ as a threshold parameter provides such an intuition, in the view that when $R_0<1$ the system is driven towards to its DFE, so that the infection does not spread in the population, while when $R_0>1$ the system is driven away from its DFE, so that the disease spreads exponentially \citep{delamater2019complexity,diekmann1990definition}.~In the case of an outbreak, such as COVID-19 pandemics, in which early predictions showed $R_0 \approx 2-3$ \cite{liu2020reproductive}, various NPIs are employed during the evolution of the outbreak, aiming to ``flatten" the curve of the epidemics.~The influence of the NPIs is frequently assessed by the reduction that they introduce to the effective reproduction number $R_t$, \citep{hellewell2020feasibility,kucharski2020early,pan2020association,cowling2020impact,tang2020estimation,tang2020updated}; ideally making $R_t<1$, which indicates that the disease spread will eventually die out.~In this work, the relation of the effective reproduction number $R_t$ with the timescales characterizing the evolution of the epidemic spread was examined in the case of COVID-19 pandemics in Italy from February 26 to September 30.

In particular, it was demonstrated analytically on the basis of the SIRD model and numerically on the basis of the SEIRD model in Section~\ref{ss:SIRD+SEIRD}, that when two of the timescales characterizing the evolution of the epidemic spread are of explosive nature, the effective reproduction number is above its threshold value; i.e., $R_t>1$.~On the contrary, when all the timescales are of dissipative nature it is implied that $R_t<1$.~In addition, the following trending behaviour was revealed: as the gap between the two explosive timescales increases/decreases, $R_t$ approaches to/withdraws from its threshold value 1, as shown in Fig.~\ref{fig:Tmscls+R0_SEIRD}.~These outcomes suggest that the insights provided by the utilization of $R_t$ as a threshold parameter can be also obtained by timescale analysis.

This work additionally suggests that timescale analysis is a robust methodology to assess the progression of the epidemic spread, since it is not hindered by the complexity of the selected model, nor the calibration process followed to fit the model against the reported data.~Following the same model calibration procedure to the SEInsRD model, resulted in timescales that are almost equal to the ones of the SEIRD model; see Figs.~\ref{fig:Tmscls+R0_SEIRD} and \ref{fig:Tmscls+R0_SEInsRD}.~Such a result indicates that the relation of the explosive timescales to $R_t$ is not affected by model selection, as discussed in Section~\ref{ss:rob}.~In addition, this relation is not affected by the parameter estimation process either, as demonstrated through the analysis of the SIDARTHE model, the calibration of which in \citep{giordano2020modelling} had significant differences with the one followed here for SEIRD and SEInsRD models; see Section~\ref{ss:calib}.

In conclusion, timescale analysis is a rigorous mathematical methodology to assess the progression of an epidemic spread, since it can effectively provide the insight obtained by the reproduction number.~Timescale analysis is not hindered by model selection in contrast to the reproduction number that is highly dependable on the structure of the selected model \citep{delamater2019complexity}.~In addition, the expression of the reproduction number becomes more complex as the detail of the model increases, as shown in Eq.~\eqref{eq:Rt}; compare for example $R_t$ fo SEIRD and SIDARTHE models.~In contrast, timescale analysis can be performed in an algorithmic fashion, utilizing the diagnostic tools of \textit{Computational Singular Perturbation} \cite{lam1989understanding,lam1994csp} that have been effectively employed to address the dynamical properties of systems arising from a wide variety of fields \cite{manias2016mechanism,tingas2015autoignition,kourdis2010physical,patsatzis2019new,patsatzis2016asymptotic}.~More importantly, the use of timescale analysis for the assessment of various NPIs is promising, since it can determine via its algorithmic tools the factors that play the most significant role on the control of ongoing COVID-19 outbreak.

\section{Acknowledgements}
This publication is based upon work supported by the Khalifa University of Science and Technology, under Award No. CPRA-2020-Goussis.

%


%

%

\bibliography{sample}

\appendix

%
%
\setcounter{equation}{0}
\setcounter{table}{0}
\setcounter{figure}{0}
\section{Derivation of the effective reproduction number}
\label{app:Rt}

The  Next Generation Matrix (NGM) approach is utilized for the calculation of the basic reproduction number $R_0$  \citep{heffernan2005,van2017,van2002}.~Given a system of ODEs in the form of Eq.~\eqref{eq:VF}, let $y_j$ be the $j=1,\ldots,m$ infected population groups among all the $y_i$ populations groups of the $i=1,\ldots,n$ compartments in $\mathbf{y}$.~In turn, let $F_i(\mathbf{y})$ be the rate of appearance of new infections in the $i$-th compartment and $V_i(\mathbf{y})=V_i^-(\mathbf{y}) - V_i^+(\mathbf{y})$ the transition rates out of ($V^-$) and into ($V^+$) the $i$-th compartment.~By definition, it is implied that:
\begin{equation}
\dfrac{dy_i}{dt}=F_i(\mathbf{y})-V_i(\mathbf{y})=F_i(\mathbf{y})+V^+_i(\mathbf{y})-V^-_i(\mathbf{y})
\label{eq:NG1}
\end{equation}
Let the matrices $\mathbf{F}$ and $\mathbf{V}$ be:
\begin{equation}
\mathbf{F}=\left[ \dfrac{\partial F_i (\mathbf{y^*})}{\partial y_j} \right] \qquad \text{and} \qquad \mathbf{V}=\left[ \dfrac{\partial V_i (\mathbf{y^*})}{\partial y_j} \right]
\label{eq:NG2}
\end{equation}
where $\mathbf{y^*}$ is the disease-free equilibrium and $i,j=1,\ldots,m$.~According to the NGM approach, the basic reproduction number $R_0$ is the spectral radius (largest eigenvalue) of the matrix $\mathbf{F}\cdot \mathbf{V^{-1}}$; i.e., $R_0=\rho(\mathbf{F}\cdot \mathbf{V^{-1}})$ \citep{heffernan2005,van2017,van2002}.~However, since the model parameters vary in time (different parameter values in each week), the NGM approach utilization results in the calculation of the effective reproduction number $R_t$.~In the following, the analytical expressions of $R_t$ for SIRD, SEIRD, SEInsRD and SIDARTHE models in Eq.~\eqref{eq:Rt} are derived.

The SIRD mathematical model in Eq.~\eqref{eq:SIRD} can be written in the form of Eq.~\eqref{eq:NG1} as:
\begin{equation}
\dfrac{d}{dt} \begin{bmatrix} S \\ I \\ R \\ D \end{bmatrix}=  \begin{bmatrix} 0 \\ \beta SI \\ 0 \\ 0 \end{bmatrix} + \begin{bmatrix} 0 \\  0 \\ \gamma I \\ \mu I \end{bmatrix} - \begin{bmatrix} \beta SI  \\  (\gamma + \mu) I \\  0 \\ 0 \end{bmatrix} = F_i(\mathbf{y})+V^+_i(\mathbf{y})-V^-_i(\mathbf{y})
\label{eq:SIRD1}
\end{equation}
The disease-free equilibrium is $\mathbf{y^*}=(S(0),0,0,0)$, so that substitution in Eq.~\eqref{eq:NG2} leads to:
\begin{equation}
\mathbf{F}=\beta S(0) \qquad \text{and} \qquad 
\mathbf{V}=\gamma + \mu
\label{eq:SIRD2}
\end{equation}
Given that $S(0)=1$ as fraction of the total population, the effective reproduction number for the SIRD model is:
\begin{equation}
R_t=\rho(\mathbf{F}\cdot \mathbf{V^{-1}})= \dfrac{\beta}{\gamma+\mu}
\label{eq:SIRD3}
\end{equation}


The SEIRD mathematical model in Eq.~\eqref{eq:SEIRD} can be written in the form of Eq.~\eqref{eq:NG1} as:
\begin{equation}
\dfrac{d}{dt} \begin{bmatrix} S \\ E \\ I \\ R \\ D \end{bmatrix}=
\begin{bmatrix} 0 \\  \beta SI \\ 0  \\ 0 \\ 0 \end{bmatrix} +
\begin{bmatrix} 0 \\  0 \\ \sigma E \\ \gamma I  \\ \mu I \end{bmatrix} -
\begin{bmatrix} \beta SI \\  \sigma E \\  \gamma I + \mu I  \\ 0 \\ 0 \end{bmatrix}
 = F_i(\mathbf{y})+V^+_i(\mathbf{y})-V^-_i(\mathbf{y})
\label{eq:SEIRD1}
\end{equation}
The disease-free equilibrium is $\mathbf{y^*}=(S(0),0,0,0,0)$, so that substitution in Eq.~\eqref{eq:NG2} leads to:
\begin{equation}
\mathbf{F}=\begin{bmatrix} 0 & \beta S(0) \\ 0 & 0 \end{bmatrix} \qquad \text{and} \qquad 
\mathbf{V}=\begin{bmatrix} \sigma & 0 \\ -\sigma & \gamma+\mu \end{bmatrix}
\label{eq:SEIRD2}
\end{equation}
Given that $S(0)=1$, the effective reproduction number for the SEIRD model is:
\begin{equation}
R_t=\rho(\mathbf{F}\cdot \mathbf{V^{-1}})=\dfrac{\beta}{\gamma+\mu} 
\label{eq:SEIRD3}
\end{equation}
Note that the $R_t$ of SEIRD model is the same to that of SIRD model in Eq.~\eqref{eq:SIRD3}.


The SEInsRD mathematical model in Eq.~\eqref{eq:SEInsRD} can be written in the form of Eq.~\eqref{eq:NG1} as:
\begin{equation}
\dfrac{d}{dt} \begin{bmatrix} S \\ E \\ IN \\ IS \\ R \\ D \end{bmatrix}= 
\begin{bmatrix} 0 \\  \beta_N S.IN + \beta_S S.IS  \\ 0  \\ 0 \\ 0 \\0 \end{bmatrix} + \begin{bmatrix} 0 \\  0 \\ (1-ss) \sigma E \\ ss \sigma E \\ \gamma (IN+IS)  \\ \mu_N IN + \mu_S IS \end{bmatrix} 
- \begin{bmatrix} \beta_N S.IN + \beta_S S.IS + \mu_{TP} S \\  \sigma E + \mu_{TP} E \\  \gamma IN + \mu_N IN  \\ \gamma IS + \mu_S IS \\  \mu_{TP} R \\ 0 \end{bmatrix}
 = F_i(\mathbf{y})+V^+_i(\mathbf{y})-V^-_i(\mathbf{y})
\label{eq:SEInsRD1}
\end{equation}
The disease-free equilibrium is $\mathbf{y^*}=(S(0),0,0,0,0,0)$, so that substitution in Eq.~\eqref{eq:NG2} leads to:
\begin{equation}
\mathbf{F}=\begin{bmatrix} 0 & \beta_N S(0) & \beta_S S(0) & \\ 0 & 0 & 0 \\ 0 & 0 & 0 \end{bmatrix} \qquad \text{and} \qquad 
\mathbf{V}=\begin{bmatrix} \sigma+\mu_{TP} & 0 & 0 \\ -(1-ss)\sigma & \gamma+\mu_N & 0  \\ - ss \sigma & 0 & \gamma+\mu_S \end{bmatrix}
\label{eq:SEInsRD2}
\end{equation}
Given that $S(0)=1$, the effective reproduction number for the SEInsRD model is:
\begin{equation}
R_t=\rho(\mathbf{F}\cdot \mathbf{V^{-1}})=\dfrac{\sigma}{\sigma+\mu} \left( \dfrac{(1-ss) \beta_N}{\gamma+\mu_N} +\dfrac{ss \beta_S}{\gamma+\mu_S}\right) 
\label{eq:SEInsRD3}
\end{equation}
Note that when considering the $\mu_{TP} \ll \sigma$ limit, $R_t$ of SEInsRD model in Eq.~\eqref{eq:SEInsRD3} is simplified to:
\begin{equation}
R_t\stackrel{\mu_{TP} \ll \sigma}{=} \left( \dfrac{(1-ss) \beta_N}{\gamma+\mu_N} +\dfrac{ss \beta_S}{\gamma+\mu_S}\right)
\end{equation}
which is similar to that of SIRD and SEIRD models in Eqs.~(\ref{eq:SIRD3}, \ref{eq:SEIRD3}) when setting $ss=0$; i.e., when neglecting the severely infected individuals from the model.


Finally, the SIDARTHE mathematical model in \citep{giordano2020modelling} can be written in the form of Eq.~\eqref{eq:NG1} as:
\begin{equation*}
\dfrac{d}{dt} \begin{bmatrix} S \\ I \\ D \\ A \\ R \\ T \\ H \\ E \end{bmatrix}= \begin{bmatrix} - S ( \alpha I - \beta D - \gamma A - \delta R) \\  S ( \alpha I + \beta D + \gamma A + \delta R) - (\epsilon+\zeta+\lambda) I \\ \epsilon I -(\eta+\rho) D \\ \zeta I -(\theta+\mu+\kappa) A \\ \eta D + \theta A - (\nu+\xi) R \\ \mu A + \nu R - (\sigma+\tau) T \\ \lambda I + \rho D + \kappa A + \xi R + \sigma T \\ \tau T \end{bmatrix} =\begin{bmatrix} 0 \\  S ( \alpha I + \beta D + \gamma A + \delta R) \\ 0 \\ 0 \\ 0 \\ 0 \\ 0 \\ 0 \end{bmatrix} +
\end{equation*}
\begin{equation}
+ \begin{bmatrix}  0 \\  0 \\ \epsilon I \\ \zeta I \\ \eta D + \theta A  \\ \mu A + \nu R \\ \lambda I + \rho D + \kappa A + \xi R + \sigma T \\ \tau T \end{bmatrix}
- \begin{bmatrix} S ( \alpha I - \beta D - \gamma A - \delta R) \\  (\epsilon+\zeta+\lambda) I \\  (\eta+\rho) D \\ (\theta+\mu+\kappa) A \\ (\nu+\xi) R \\  (\sigma+\tau) T \\ 0 \\ 0 \end{bmatrix} 
= F_i(\mathbf{y})+V^+_i(\mathbf{y})-V^-_i(\mathbf{y})
\label{eq:SIDARTHE1}
\end{equation}
where the parameter notation is explained in detail in \citep{giordano2020modelling}.~The disease-free equilibrium is $\mathbf{y^*}=(S(0),0,0,0,0,0,0,0)$, so that substitution in Eq.~\eqref{eq:NG2} leads to:
\begin{equation}
\mathbf{F}=\begin{bmatrix} \alpha S(0) & \beta S(0) & \gamma S(0) & \delta S(0) & 0 \\ 0 & 0 & 0 & 0 & 0 \\ 0 & 0 & 0 & 0 & 0 \\ 0 & 0 & 0 & 0 & 0 \\ 0 & 0 & 0 & 0 & 0 \end{bmatrix} \qquad \text{and} \qquad 
\mathbf{V}=\begin{bmatrix} \epsilon+\lambda+\zeta & 0 & 0 & 0 & 0 \\ -\epsilon & \eta+\rho & 0 & 0 & 0 \\ -\zeta & 0 & \kappa+\mu+\theta & 0 & 0 \\ 0 & -\eta & -\theta & \nu+\xi & 0 \\ 0 & 0 & \mu & \nu & \sigma + \tau \end{bmatrix}
\label{eq:SIDARTHE2}
\end{equation}
Given that $S(0)=1$, the effective reproduction number for the SIDARTHE model is:
\begin{equation}
R_t=\rho(\mathbf{F}\cdot \mathbf{V^{-1}})= \dfrac{\alpha}{r_1} + \dfrac{\beta \epsilon}{r_1 r_2} + \dfrac{\gamma \zeta}{r_1 r_3} + \dfrac{\delta \theta \zeta}{r_1 r_3 r_4} + \dfrac{\delta \epsilon \eta}{r_1 r_2 r_4}
\label{eq:SIDARTHE3}
\end{equation}
where $r_1=\epsilon+\lambda+\zeta$, $r_2=\eta+\rho$, $r_3=\kappa+\mu+\theta$ and $r_4=\nu+\xi$.~Note that the expression in Eq.~\eqref{eq:SIDARTHE3} derived here in the context of NGM approach is the same with the one in Eq.~(18) derived in \citep{giordano2020modelling} using a different approach.

\section{The SEIRD and SEInsRD model parameters}
\label{app:ModParam}

\begin{figure}[!h]
\centering
\includegraphics[scale=0.23]{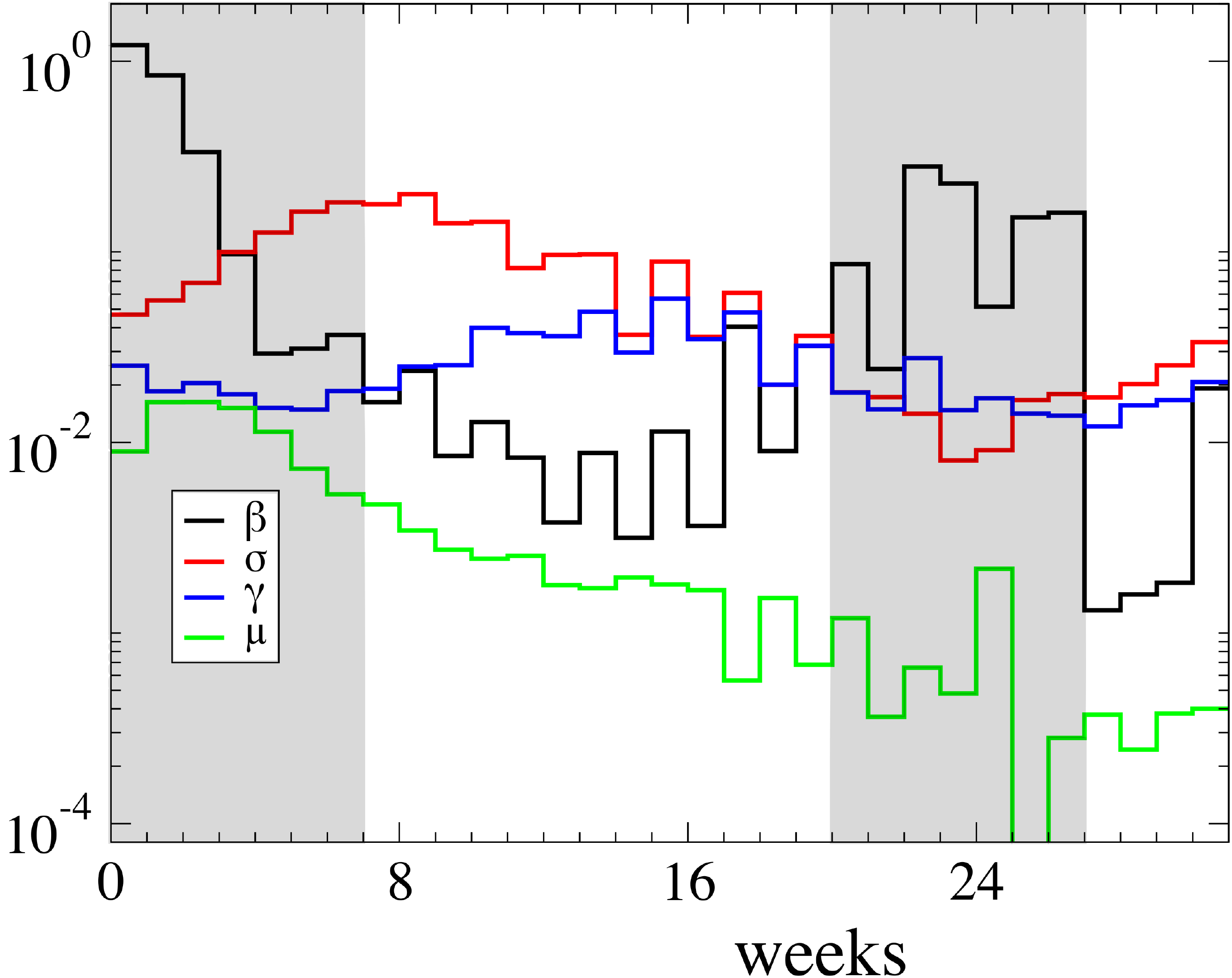} \hspace{2pt} \includegraphics[scale=0.23]{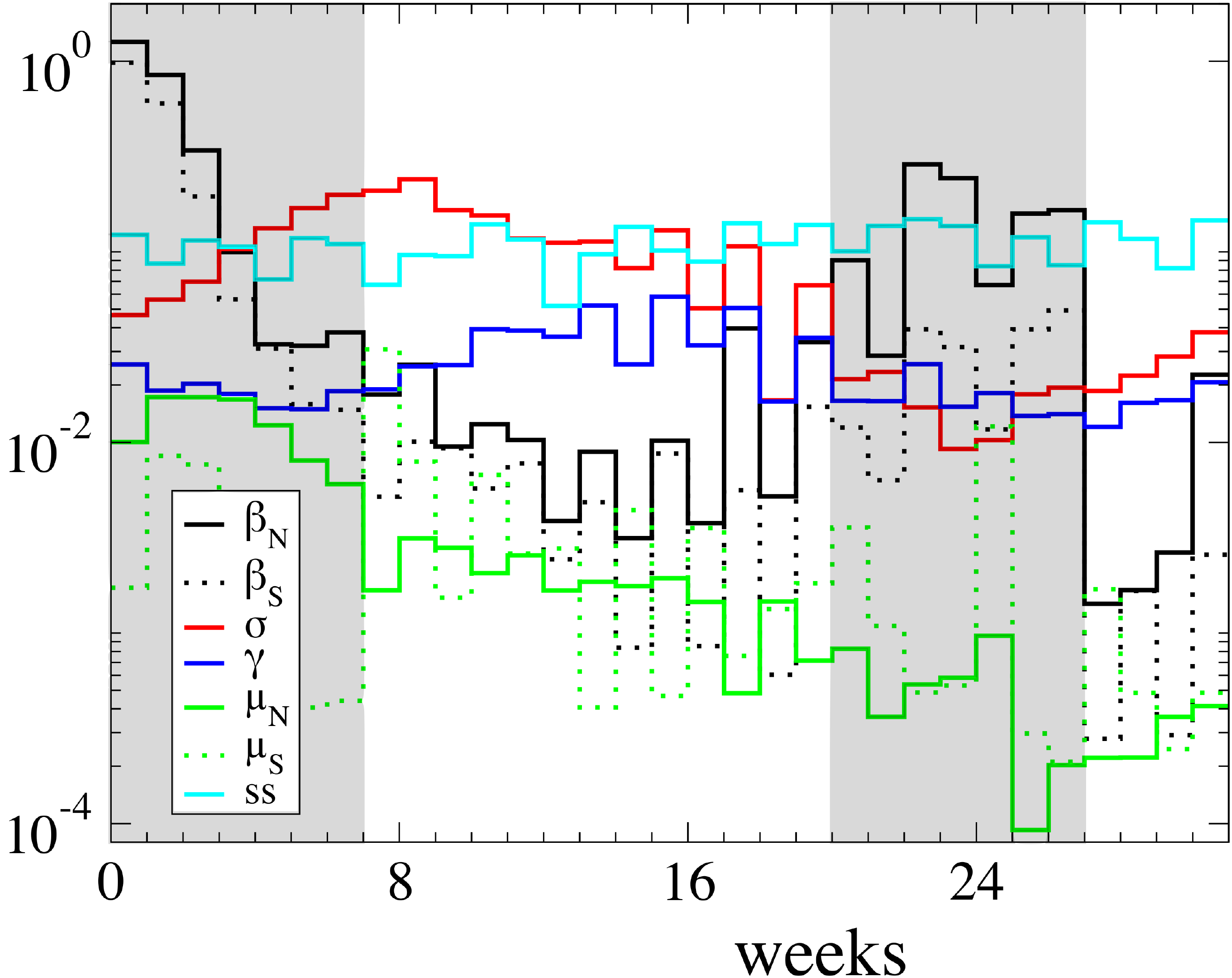}
\caption{The parameters estimated for the SEIRD (left) and SEInsRD (right) models.~The shaded regions indicate the weeks for which $R_t>1$ and explosive timescales arise.}
\label{fig:Params_SEIRD+SEInsRD}
\end{figure}

The parameter estimation process described in Section~\ref{ss:calib}, that was followed to fit the reported data sets of infected, recovered and dead individuals of Italy from February 26 to September 30 in a weekly basis, resulted in the model parameters shown in Fig.~\ref{fig:Params_SEIRD+SEInsRD}.~The left panel shows the distribution of the SEIRD model parameters $\beta$, $\sigma$, $\gamma$ and $\mu$ and the right panel shows the ones of SEInsRD model $\beta_N$, $\beta_S$, $\sigma$, $\gamma$, $\mu_N$, $\mu_S$ and $ss$.~The values of parameter $\mu_{TP}$ of the SEInsRD model are not shown, since they are smaller than $10^{-5}$.

Figure~\ref{fig:Params_SEIRD+SEInsRD} indicates that the parameters expressing the transition from a population group to another attain similar values in both models: transmission rate ($\beta$ and $\beta_N, \beta_S$), incubation period ($1/\sigma$), recovery rate ($\gamma$) and fatality rate ($\mu$ and $\mu_N, \mu_S$) constants.

As shown in Fig.~\ref{fig:Params_SEIRD+SEInsRD}, the following trends in the parameter values are indicated:
\begin{itemize}
\item the transmission rate constant $\beta$ attains high/low values in the periods where explosive timescale are present/absent.~The values of $\beta$ tend to decrease during the transition from an explosive to a dissipative region and vice-versa.
\item the rate constant $\sigma$ (inverse of incubation period) tend to increases during the explosive regions.
\item the recovery rates $\gamma$ are almost constant
\item the fatality rates $\mu$ tend to decrease, despite the explosive/dissipative region transition.~They tend to increase only in the last few weeks.
\item the normally to severely infected ratio $ss$ is almost constant.
\end{itemize}

\begin{table}[!h]
\centering
\begin{tabular}{l | l | l }
 population group & SEIRD & SEInsRD\\
\hline \hline
infected, $I$ & $0.99972$ & $0.99813$ \\
recovered, $R$ & $0.99993$ & $0.99985$ \\
dead, $D$ & $0.99998$ & $0.99969$ \\ 
\end{tabular}
\caption{$R^2$ values of the solution acquired on the basis of the SEIRD and SEInsRD models with the parameter distribution shown in Fig.~\ref{fig:Params_SEIRD+SEInsRD}, with reference to the reported data for infected, recovered and dead individuals in Italy.}
\label{tb:ParamRange+Rs+p}
\end{table}

\end{document}